\documentclass[referee]{raa}

\usepackage{graphicx,times}
\usepackage{natbib}
\usepackage{amssymb,amsmath}
\usepackage{multirow}
\usepackage{booktabs}
\bibpunct{(}{)}{;}{a}{}{,}

\usepackage[pagebackref=true]{hyperref}

\begin{document}

  \title{Relationship of ICME Composition Signatures with Solar Activity during 2009--2025}

   \volnopage{Vol.0 (202x) No.0, 000--000}
   \setcounter{page}{1}

   \author{Diyorbek Pulatov
      \inst{1}\footnotetext{$*$Corresponding Author.}
   \and Zavkiddin Mirtoshev
      \inst{1,*}
    \and Mirabbos Mirkamalov
       \inst{2}
   }

   \institute{Department of Nuclear Physics and Astronomy, Institute of Nuclear Technologies, Samarkand State University,
             Samarkand 140104, Uzbekistan; {\it zmirtoshev@samdu.uz}\\
    \and
             CAS Key Laboratory of Geospace Environment, Department of Geophysics and Planetary Sciences, University of Science and Technology of China, Hefei, Anhui 230026, People’s Republic of China\\
\vs\no
   {\small Received 202x month day; accepted 202x month day}}

\abstract{
Coronal Mass Ejections (CMEs) are among the most energetic solar eruptions, expelling magnetized plasma from the corona into interplanetary space. Their interplanetary counterparts, known as Interplanetary Coronal Mass Ejections (ICMEs), retain distinct compositional signatures that reflect the physical conditions of their solar source regions. This study presents a statistical analysis of ICME composition dependence on solar activity during 2009--2025, covering Solar Cycle (SC) 24 and the ascending phase of SC~25 through its maximum, and compares the results with SC~23 studied by \citet{Song+etal+2021}. Using data from the Solar Wind Ion Composition Spectrometer (SWICS) aboard the Advanced Composition Explorer (ACE), we examined the average iron charge state ($\langle Q_{\rm Fe}\rangle$), ionic ratios (C$^{6+}$/C$^{5+}$ and O$^{7+}$/O$^{6+}$), and the elemental abundance ratio (Fe/O) for 307 ICMEs listed in the Richardson and Cane ICME catalog. The results show strong positive correlations of $\langle Q_{\rm Fe}\rangle$ ($r$~=~0.86) and O$^{7+}$/O$^{6+}$ ($r$~=~0.85) with the annual sunspot number (SSN), whereas C$^{6+}$/C$^{5+}$ exhibits a weak correlation ($r$~=~0.17) primarily due to SWICS~2.0 upper-end saturation truncation that suppresses the solar maximum signal, and Fe/O a moderate correlation ($r$~=~0.57). The Fe/O ratio, a proxy for the First Ionization Potential (FIP) effect, displayed elevated values during the maxima of both SC~24 and SC~25, suggesting enhanced elemental fractionation during periods of increased magnetic activity. Comparing with SC~23, we find that the overall solar cycle dependence of ICME composition persists across cycles, though with notable quantitative differences attributed to the different magnetic activity levels between cycles. These findings confirm that ICME compositional signatures are strongly modulated by the solar cycle, offering insights into CME initiation, coronal plasma processes, and their implications for space weather forecasting.
}

   \authorrunning{D. Pulatov, Z. Mirtoshev \& M. Mirkamalov}
   \titlerunning{ICME Composition and Solar Activity}

   \maketitle

\section{Introduction}
\label{sect:intro}

Coronal mass ejections (CMEs) are large-scale eruptions of magnetized plasma from the solar corona into the heliosphere \citep{Webb+Howard+2012}. CMEs often exhibit a three-part structure in white-light images, including a bright front, a dark cavity, and a bright core \citep{Howard+etal+1997, Song+etal+2023}. CMEs are usually driven by the eruption and magnetohydrodynamic instability of magnetic flux ropes formed prior to \citep{Patsourakos+etal+2013} or during \citep{Song+etal+2014} solar eruptions \citep{Chen+2011, Manchester+etal+2017}. They transport large amounts of coronal magnetic field and plasma into interplanetary space, where they are detected as interplanetary coronal mass ejections (ICMEs) through in-situ measurements \citep{Zurbuchen+Richardson+2006}.

ICMEs are characterized by enhanced magnetic fields, depressed proton temperatures, bidirectional electron flows, and distinct heavy-ion composition signatures \citep{Zurbuchen+Richardson+2006}. They can drive geomagnetic storms when they impact the Earth's magnetosphere \citep{Gosling+etal+1991, Zhang+etal+2007, Gopalswamy+2009}. Magnetic clouds (MCs) are a well-defined subset of ICMEs, typically detected in approximately one-third of ICME events. MCs are characterized by enhanced magnetic field strengths ($>$10~nT), smooth large-scale field rotations, and low ion temperatures \citep{Burlaga+etal+1981}. \citet{Song+etal+2016} performed a statistical analysis of the average iron charge state distributions within 96 MCs during SC 23 and found that 11 of these events exhibited bimodal charge state distributions, indicating the coexistence of both hot and cold plasma populations within individual MCs. Furthermore, \citet{Owens+2018} compared 97 MC events with 118 non-MC events and found systematically higher ionic charge states and elemental abundances in MC events.

The ionic charge states and elemental abundances within ICMEs serve as valuable diagnostics for understanding CME source regions and eruption processes \citep{Song+Yao+2020}. In the solar corona, elements with a low first ionization potential (FIP; $<$10~eV), such as Mg, Fe, and Si, are systematically enhanced relative to high-FIP elements (O, Ne, He) compared to photospheric values. This phenomenon, known as the ``FIP effect,'' was first recognized through early X-ray and ultraviolet observations demonstrating that coronal abundances of low-FIP elements significantly exceed their photospheric values \citep{Pottasch+1963}. The FIP effect is attributed to the ponderomotive force of Alfv\'{e}n waves acting on ions in the chromosphere, which preferentially fractionates low-FIP elements into the corona \citep{Laming+2015}. The Fe/O elemental abundance ratio is widely used as a measure of the FIP effect and to differentiate the coronal source regions of solar wind and ICMEs \citep{Zurbuchen+etal+2016}.

The ionic charge states freeze-in at different heliocentric distances depending on the element and charge state involved. Carbon and oxygen charge states freeze-in close to the Sun, making the C$^{6+}$/C$^{5+}$ and O$^{7+}$/O$^{6+}$ ratios effective tracers of source region conditions and solar wind classification \citep{Zurbuchen+etal+2002, Landi+etal+2012}. The O$^{7+}$/O$^{6+}$ ratio is sensitive to the local electron temperature and is typically elevated within ICMEs \citep{Henke+etal+2001, Kasper+etal+2012}. In contrast, the average iron charge state ($\langle Q_{\rm Fe}\rangle$) freezes-in over a broader distance range (1.2--5~$R_{\odot}$) in the solar wind and even farther in ICMEs, making it a sensitive indicator of the coronal electron temperature and more suitable for analyzing CME eruption processes \citep{Lepri+etal+2001, Lepri+Zurbuchen+2004, Lepri+etal+2013, Wang+etal+2017}.

Previous studies have established that both the charge states and elemental abundances of the solar wind exhibit solar cycle dependence \citep{Lepri+etal+2013, Zhao+etal+2014, Zhao+etal+2017}. In a comprehensive statistical study, \citet{Song+etal+2021} analyzed the charge states of five elements and the relative abundances of six element pairs within 319 ICMEs during 1998--2011 (SC~23). They found that all ICME compositions exhibit solar cycle dependence, with nearly all ionic charge states and elemental abundances (except C/O) positively correlated with the sunspot number (SSN). Notably, the Ne/O ratios of ICMEs and slow solar wind showed opposite solar cycle trends.

The purpose of this study is to extend the analysis of \citet{Song+etal+2021} to the subsequent period, examining the dependence of ICME composition on solar activity during 2009--2025. This period covers the complete SC~24 and the ascending phase and maximum of SC~25, enabling a comparison with SC~23 and providing insights into how composition varies between solar cycles of different strengths. We describe the observational data and analysis methodology in Section~\ref{sect:data}, present the results in Section~\ref{sect:results}, provide a discussion in Section~\ref{sect:discussion}, and summarize the conclusions in Section~\ref{sect:conclusion}.

\section{Observational Data and Analysis Methodology}
\label{sect:data}

\subsection{Instrumentation and Data Sources}

The compositional data used in this study are obtained from the Solar Wind Ion Composition Spectrometer (SWICS; \citealt{Gloeckler+etal+1998}) aboard the Advanced Composition Explorer (ACE), which orbits the Sun--Earth L1 Lagrangian point. The SWICS instrument identifies solar wind ions through a combination of three independent measurements: electrostatic selection of the energy-per-charge (E/q), time-of-flight velocity measurement, and total kinetic energy measurement \citep{Gilbert+etal+2012}. We obtained two-hourly averaged data for the average iron charge state ($\langle Q_{\rm Fe}\rangle$), ionic ratios (O$^{7+}$/O$^{6+}$, C$^{6+}$/C$^{5+}$), and elemental abundance (Fe/O) from the Coordinated Data Analysis Web\footnote{\url{https://cdaweb.gsfc.nasa.gov/}}.

Two versions of the SWICS data are used: SWICS~1.1 (January 2009 to August 2011) and SWICS~2.0 (June 2012 onward). A radiation and age-induced hardware anomaly on August~23, 2011 altered the operational state of ACE/SWICS, degrading its performance and introducing saturation artifacts in the post-anomaly data (SWICS~2.0). The SWICS~2.0 data were re-calibrated using statistical methods described by \citet{Shearer+etal+2014}. Following the recommendations of the ACE/SWICS instrument team \citep{Zhao+etal+2024}, we applied the following saturation filters to ensure cross-calibration between the SWICS~1.1 and 2.0 datasets.
 
For C$^{6+}$/C$^{5+}$, saturation occurs at the upper end of the distribution: all two-hourly data points with C$^{6+}$/C$^{5+}$ $\geq$~1.486 were identified using the instrument quality flags and removed from both datasets, affecting approximately 4.0\% of the SWICS~2.0 measurements (1,657 out of 41,425 data points). For O$^{7+}$/O$^{6+}$, saturation occurs at the lower end: values $\leq$~0.0523 were removed, affecting 15.8\% of the data (6,572 data points). For Fe/O, saturation at the upper end ($\geq$~0.4327) was removed, affecting only 1.2\% (478 data points). Additionally, He$^{2+}$ contamination of Fe$^{6+}$ and Fe$^{7+}$ charge states introduces noise in the SWICS~2.0 data that affects both $\langle Q_{\rm Fe}\rangle$ and Fe/O; these contaminated charge states were therefore removed from both the SWICS~1.1 and 2.0 datasets to ensure continuity across the instrument transition. For $\langle Q_{\rm Fe}\rangle$, the SWICS~2.0 data do not exhibit a simple saturation threshold as the ionic ratios do. Instead, $\langle Q_{\rm Fe}\rangle$ is affected by He$^{2+}$ contamination of the Fe$^{6+}$ and Fe$^{7+}$ charge state channels in the SWICS~2.0 data, which artificially enhances the counts in these low charge states and biases $\langle Q_{\rm Fe}\rangle$ downward. Following the instrument team recommendations \citep{Zhao+etal+2024}, we removed the Fe$^{6+}$ and Fe$^{7+}$ charge states from the $\langle Q_{\rm Fe}\rangle$ calculation in both the SWICS~1.1 and 2.0 datasets, so that $\langle Q_{\rm Fe}\rangle$ is computed from Fe$^{8+}$ through Fe$^{16+}$ only. This ensures continuity across the instrument transition and eliminates the He$^{2+}$ contamination bias. The saturation thresholds and statistics are summarized in Table~\ref{TabA1}.

 \begin{table}[h]
\bc
\begin{minipage}[]{120mm}
\caption[]{Saturation statistics for ACE/SWICS~2.0 data (June 2012 onward), based on the instrument team recommendations \citep{Zhao+etal+2024}.}\label{TabA1}
\end{minipage}
\setlength{\tabcolsep}{4pt}
\small
\begin{tabular}{lccc}
\toprule
Parameter & Saturation Threshold & No. Saturated Points & Fraction (\%) \\
\midrule
C$^{6+}$/C$^{5+}$ & $\geq$~1.48630 (upper end) & 1,657 & 4.0 \\
O$^{7+}$/O$^{6+}$ & $\leq$~0.052272 (lower end) & 6,572 & 15.8 \\
Fe/O & $\geq$~0.4327 (upper end) & 478 & 1.2 \\
\bottomrule
\end{tabular}
\ec
\end{table}

As an illustrative example, we describe the data processing for the ICME event of 2014 September 12--13 (associated with an X1.6 flare and halo CME). The two-hourly C$^{6+}$/C$^{5+}$ time series within the ICME boundaries contained 36 data points, of which 4 reached the saturation threshold ($\geq$~1.486) and were excluded. The remaining 32 data points were averaged to yield the event mean C$^{6+}$/C$^{5+}$~=~1.21. For O$^{7+}$/O$^{6+}$, none of the 36 data points fell below the 0.0523 threshold, so all values were retained, yielding a mean of 0.89. For Fe/O, 2 data points exceeded 0.4327 and were removed, yielding a mean of 0.38. The same procedure was applied to all compositional parameters for each of the 307 ICME events.

The annual average SSN data were obtained from the Solar Influences Data Analysis Center (SIDC) of the Royal Observatory of Belgium\footnote{\url{https://www.sidc.be/}} \citep{Clette+Lefevre+2016}.

\subsection{ICME Catalog}

The ICME event boundaries are from the Richardson and Cane (RC) catalog \citep{Richardson+Cane+2010, Richardson+Cane+2024}. During January 2009 to August 2025, a total of 307 ICMEs are listed in the RC catalog.

\begin{figure}
\centering
\includegraphics[width=0.7\textwidth]{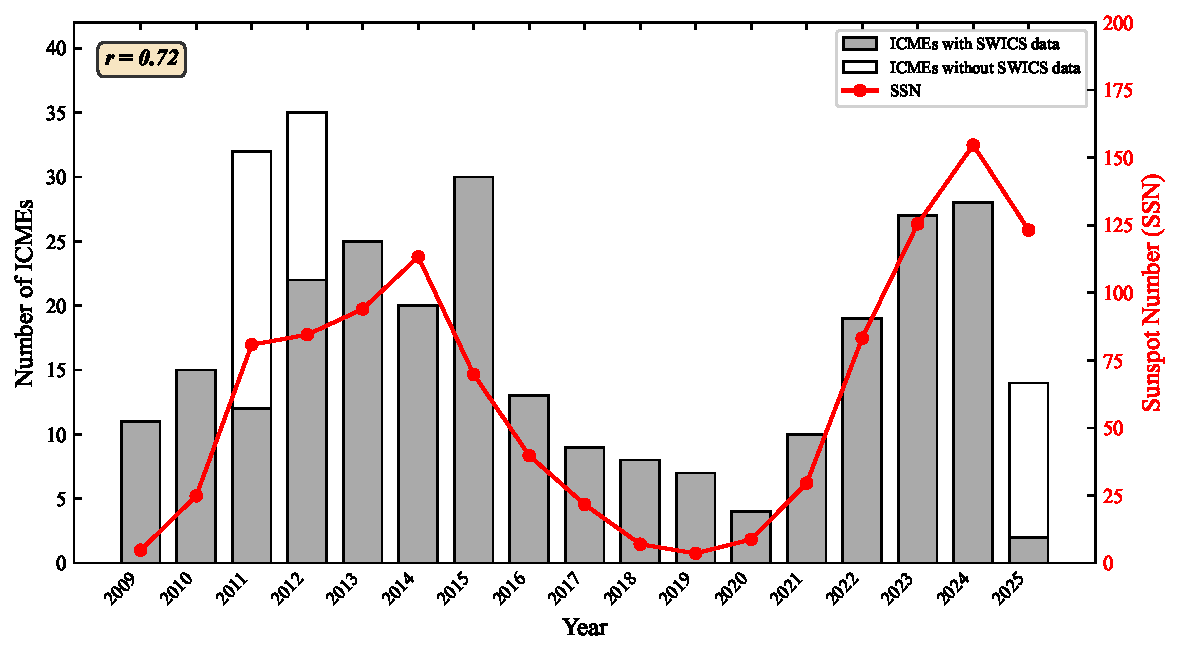}
\caption{Annual number of ICMEs (histogram) and SSN (red line with filled circles) from 2009 to 2025 during SC~24 and SC~25. Gray-shaded portions represent ICMEs with available SWICS data; white portions indicate ICMEs without data. The Pearson coefficient between ICME counts and SSN is displayed.}
\label{Fig1}
\end{figure}

Figure~\ref{Fig1} displays the annual ICME numbers from 2009 to 2025 as histograms, together with the yearly averaged SSN shown as a red line. The gray-shaded portions indicate ICMEs with available SWICS data (262 events), while the white portions represent ICMEs without available data. Notable data gaps exist in 2011 (20 of 32 events without data), 2012 (13 of 35), and 2025 (12 of 14). We computed Pearson ($r$), Spearman ($\rho$), and Kendall ($\tau$) correlation coefficients to assess linear, monotonic, and rank-based relationships. The Pearson coefficient between the annual ICME number and SSN is $r$~=~0.72, confirming a strong positive relationship consistent with SC~23 ($r$~=~0.85; \citealt{Song+etal+2021}), while the Spearman ($\rho$ = 0.74) and Kendall ($\tau$ $\approx$ 0.53) coefficients indicate a statistically significant positive relationship between ICME occurrence rate and solar activity.

\subsection{Analysis Methodology}

For each ICME, the average value of each compositional parameter was computed from the two-hourly SWICS data within the ICME boundaries defined by the RC catalog. The annual mean and standard deviation of each parameter were then calculated from all ICME events in a given year. The standard deviation reflects event-to-event variability, which can be substantial due to differences in source regions, eruption processes, and spacecraft trajectories \citep{Song+Yao+2020}. To quantify the relationship between composition and solar activity, we computed the correlation coefficients between yearly means and SSN.

Three complementary correlation coefficients are used. The Pearson coefficient ($r$) measures the strength of the linear relationship and is sensitive to outliers. The Spearman rank coefficient ($\rho$) evaluates the monotonic relationship by ranking the data, making it robust against outliers and non-normal distributions; a higher $\rho$ than $r$ indicates a monotonic but nonlinear dependence. The Kendall rank coefficient ($\tau$) counts concordant and discordant pairs, providing the most conservative and robust estimate, particularly suitable for small sample sizes. For a dataset of $N$~=~17 yearly values, we require $|r|$~$>$~0.48 for significance at the $p < 0.05$ level; thus correlations with $|r|$~$<$~0.48 are not statistically significant.

\section{Results}
\label{sect:results}

\subsection{Solar Cycle Dependence of ICME Composition Parameters}
\label{sect:results_SC}

\begin{figure}
\centering
\includegraphics[width=0.7\textwidth]{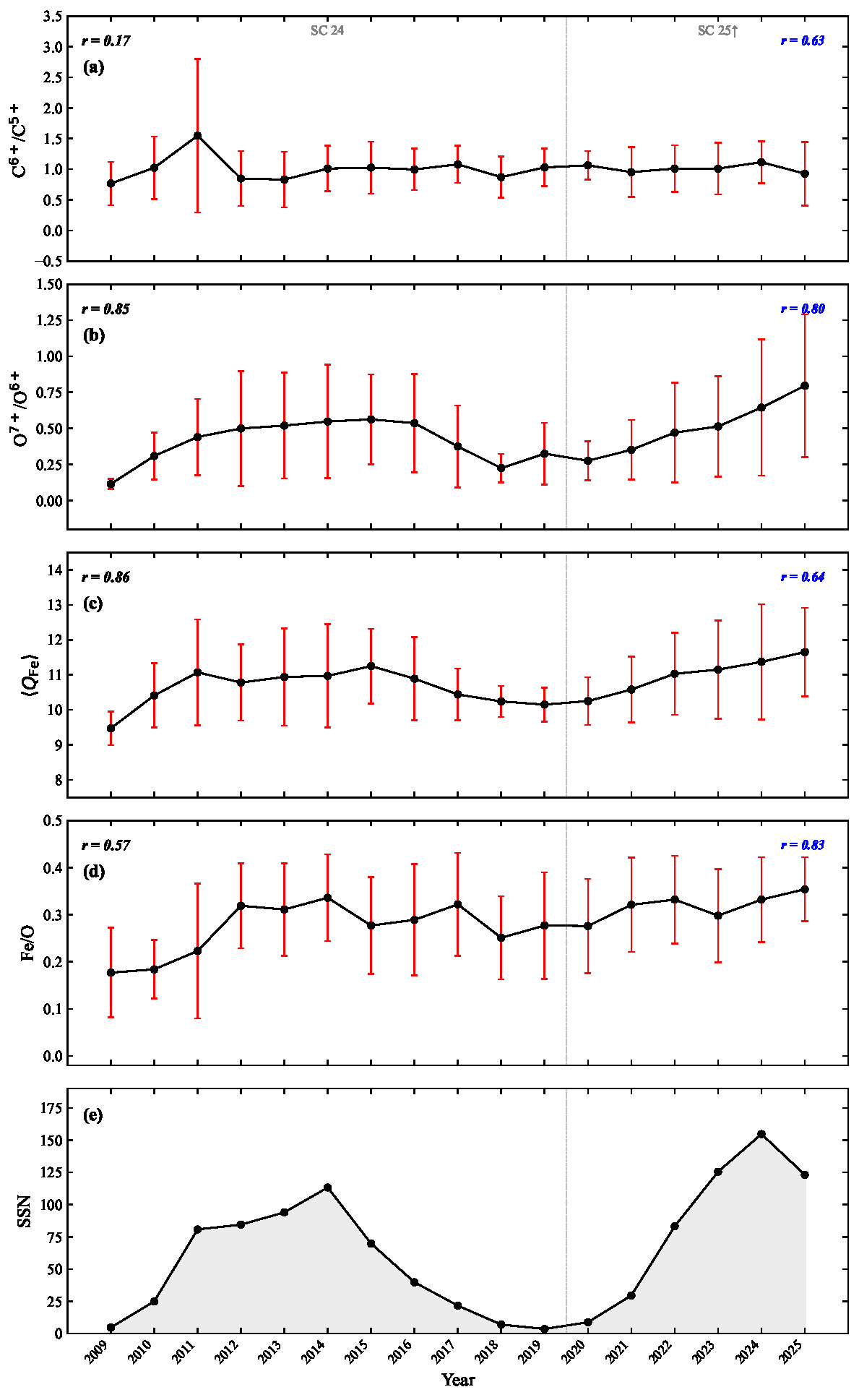}
\caption{Solar cycle dependence of ICME composition parameters from 2009 to 2025. Panels show annual means and standard deviations (red bars) of (a)~C$^{6+}$/C$^{5+}$, (b)~O$^{7+}$/O$^{6+}$, (c)~$\langle Q_{\rm Fe}\rangle$, (d)~Fe/O, and (e)~SSN. Black $r$ values are from this study (2009--2025); blue italic values are from \citet{Song+etal+2021} (SC~23). The dashed vertical line marks the SC~24/25 boundary.}
\label{Fig2}
\end{figure}

Table~\ref{Tab1} presents the yearly mean values and standard deviations of the four compositional parameters from 2009 to 2025. Figure~\ref{Fig2} displays the temporal evolution of these parameters compared with the SSN, showing annual means and standard deviations (vertical red bars) for (a)~C$^{6+}$/C$^{5+}$, (b)~O$^{7+}$/O$^{6+}$, (c)~$\langle Q_{\rm Fe}\rangle$, and (d)~Fe/O, with panel~(e) showing the SSN. The Pearson coefficients from the present study (black) and from \citet{Song+etal+2021} for SC~23 (blue, italic) are indicated in each panel. The full results of the correlation analysis are summarized in Table~\ref{tab:correlations}.

\begin{table}
\bc
\begin{minipage}[]{160mm}
\caption[]{Annual Mean Values and Standard Deviations of ICME Compositional Parameters from 2009 to 2025}\label{Tab1}
\end{minipage}
\setlength{\tabcolsep}{2pt}
\small
\begin{tabular}{lcccccccccccccccccc}
\toprule
Parameter & & 2009 & 2010 & 2011 & 2012 & 2013 & 2014 & 2015 & 2016 & 2017 & 2018 & 2019 & 2020 & 2021 & 2022 & 2023 & 2024 & 2025\\
\midrule
No. ICMEs & & 11 & 15 & 12 & 22 & 25 & 20 & 30 & 13 & 9 & 8 & 7 & 4 & 10 & 19 & 27 & 28 & 2\\
\midrule
    C$^{6+}$/C$^{5+}$ & mean & 0.77 & 1.02 & 1.55 & 0.85 & 0.83 & 1.01 & 1.03 & 1.00 & 1.08 & 0.87 & 1.03 & 1.06 & 0.95 & 1.01 & 1.01 & 1.12 & 0.93\\
 & $\sigma$ & 0.35 & 0.51 & 1.26 & 0.45 & 0.46 & 0.37 & 0.42 & 0.34 & 0.30 & 0.33 & 0.31 & 0.23 & 0.41 & 0.38 & 0.42 & 0.34 & 0.52\\
\midrule
O$^{7+}$/O$^{6+}$ & mean & 0.11 & 0.31 & 0.44 & 0.50 & 0.52 & 0.55 & 0.56 & 0.54 & 0.37 & 0.22 & 0.32 & 0.28 & 0.35 & 0.47 & 0.51 & 0.64 & 0.80\\
 & $\sigma$ & 0.04 & 0.16 & 0.26 & 0.40 & 0.37 & 0.39 & 0.31 & 0.34 & 0.28 & 0.10 & 0.22 & 0.13 & 0.21 & 0.35 & 0.35 & 0.47 & 0.50\\
\midrule
$\langle Q_{\rm Fe}\rangle$ & mean & 9.47 & 10.41 & 11.07 & 10.78 & 10.94 & 10.97 & 11.25 & 10.89 & 10.44 & 10.24 & 10.15 & 10.25 & 10.58 & 11.03 & 11.15 & 11.37 & 11.65\\
 & $\sigma$ & 0.48 & 0.92 & 1.52 & 1.09 & 1.39 & 1.48 & 1.07 & 1.19 & 0.74 & 0.45 & 0.49 & 0.68 & 0.94 & 1.17 & 1.40 & 1.65 & 1.27\\
\midrule
Fe/O & mean & 0.18 & 0.18 & 0.22 & 0.32 & 0.31 & 0.34 & 0.28 & 0.29 & 0.32 & 0.25 & 0.28 & 0.28 & 0.32 & 0.33 & 0.30 & 0.33 & 0.35\\
 & $\sigma$ & 0.10 & 0.06 & 0.14 & 0.09 & 0.10 & 0.09 & 0.10 & 0.12 & 0.11 & 0.09 & 0.11 & 0.10 & 0.10 & 0.09 & 0.10 & 0.09 & 0.07\\
\midrule
SSN & & 4.8 & 24.9 & 80.8 & 84.5 & 94.0 & 113.3 & 69.8 & 39.8 & 21.7 & 7.0 & 3.6 & 8.8 & 29.6 & 83.2 & 125.5 & 154.7 & 123.2\\
\bottomrule
\end{tabular}
\ec
\end{table}

% Requires: \usepackage{amsmath}
\begin{table}[ht]
    \centering
    \caption{Correlation coefficients between ICME Compositional Parameters and SSN.}
    \label{tab:correlations}
    \begin{tabular}{lcccc}
        \hline
        Parameter & Pearson $r$ & Spearman $\rho$ & Kendall $\tau$ & Interpretation \\ \hline
        C$^{6+}$/C$^{5+}$ & 0.17 & 0.01 & 0.01 & No statistically significant relationship \\
        O$^{7+}$/O$^{6+}$ & 0.85 & 0.84 & 0.67 & Strongly monotonic and linear \\
        $\langle Q_{\rm Fe}\rangle$ & 0.86 & 0.89 & 0.69 & Strongly monotonic, slightly nonlinear \\
        Fe/O & 0.57 & 0.67 & 0.49 & Moderate monotonic, nonlinear \\ \hline
    \end{tabular}
\end{table}

\textit{Carbon ionic ratio (C$^{6+}$/C$^{5+}$).---}Panel~(a) of Figure~\ref{Fig2} shows that C$^{6+}$/C$^{5+}$ exhibits a weak dependence on SSN. The coefficients ($r$ = 0.17, $\rho$ = 0.01, $\tau$ = 0.01) indicate no statistically significant relationship, implying non-monotonic, multi-regime variability driven by local coronal conditions. From the SC~23/24 minimum, C$^{6+}$/C$^{5+}$ increased from 0.77 (2009) to 1.55 in 2011, then decreased sharply to 0.85 (2012) and 0.83 (2013) before recovering to 1.01--1.03 during 2014--2015. During SC~25, the values remained near 0.95--1.12 without a clear monotonic trend mirroring the SSN rise. The $r$ value (0.17) is substantially lower than reported for SC~23 ($r$ = 0.63; \citealt{Song+etal+2021}). The C$^{6+}$/C$^{5+}$ ratio freezes-in in the lower solar corona and is used to classify solar wind types \citep{Zhao+etal+2009}. The weaker correlation during SC~24/25 is primarily attributable to the SWICS~2.0 upper-end saturation truncation (see below). A secondary contributing factor may be that C$^{6+}$/C$^{5+}$ freezes-in at lower heliocentric distances ($<$1.5~$R_{\odot}$) than $\langle Q_{\rm Fe}\rangle$ or O$^{7+}$/O$^{6+}$, making it more sensitive to conditions in the low corona and to potential contamination from surrounding solar wind at imprecise ICME boundaries, rather than to the global coronal temperature that tracks the solar cycle.

The substantially weaker correlation of C$^{6+}$/C$^{5+}$ with SSN ($r$~=~0.17) compared to SC~23 ($r$~=~0.63) can be attributed primarily to an instrumental artifact in the SWICS~2.0 data. The C$^{6+}$/C$^{5+}$ ratio saturates at the upper end of the distribution, with all values $\geq$~1.486 removed (Section~2.1). This truncation preferentially suppresses the high C$^{6+}$/C$^{5+}$ values that are most frequently produced in flare-associated ICMEs during solar maximum. During SC~23, \citet{Song+etal+2021} reported yearly mean C$^{6+}$/C$^{5+}$ values reaching 1.66--2.17 during the solar maximum years (2001--2004), well above the 1.486 saturation threshold. In the present study, the only year with C$^{6+}$/C$^{5+}$ exceeding 1.486 (1.55 in 2011) falls within the pre-anomaly SWICS~1.1 period. For the entire SWICS~2.0 period (2012--2025), yearly means range from only 0.83 to 1.12, a compressed dynamic range of $\sim$0.29 compared to $\sim$1.45 during SC~23. This strongly suppressed variability eliminates the solar cycle signal. The Spearman ($\rho$~=~0.01) and Kendall ($\tau$~=~0.01) coefficients further confirm the absence of any monotonic trend in the truncated data. Consequently, the low $r$ value is primarily a data limitation rather than evidence that ICME C$^{6+}$/C$^{5+}$ lacks solar cycle dependence. Future composition measurements from Solar Orbiter and Parker Solar Probe will be essential to verify the true C$^{6+}$/C$^{5+}$ behavior across the solar cycle.

\textit{Oxygen ionic ratio (O$^{7+}$/O$^{6+}$).---} Panel~(b) of Figure~\ref{Fig2} demonstrates that O$^{7+}$/O$^{6+}$ tracks the SSN variation closely. The coefficients ($r$ = 0.85, $\rho$ = 0.84, $\tau$ = 0.67) indicate a strongly linear and monotonic relationship, showing that oxygen charge states scale with solar activity. The high $\tau$ confirms consistent ordering across the cycle, supporting O$^{7+}$/O$^{6+}$ as a reliable tracer of large-scale coronal temperature variations. O$^{7+}$/O$^{6+}$ rose from 0.11 in 2009 to 0.55--0.56 during SC~24 maximum (2014--2015), then decreased to 0.22 in 2018 before recovering during SC~25 to 0.64 (2024) and reaching 0.80 in 2025. For comparison, \citet{Song+etal+2021} reported $r$~=~0.80 for SC~23. This confirms that O$^{7+}$/O$^{6+}$ is a robust tracer of solar activity.

\textit{Average iron charge state ($\langle Q_{\rm Fe}\rangle$).---}Panel~(c) of Figure~\ref{Fig2} shows that $\langle Q_{\rm Fe}\rangle$ follows the SSN closely, with a very strong positive correlation ($r$ = 0.86, $\rho$ = 0.89, $\tau$ = 0.69). The slightly higher Spearman coefficient indicates a predominantly monotonic, mildly nonlinear dependence, likely reflecting coronal heating effects such as saturation or cycle variability. This confirms that $\langle Q_{\rm Fe}\rangle$ is a robust tracer of coronal electron temperature and global magnetic activity. During the ascending phase of SC~24, $\langle Q_{\rm Fe}\rangle$ increased from 9.47 (2009) to a peak of 11.25 in 2015. It subsequently decreased during the descending phase to 10.15 in 2019 (SC~24/25 minimum). During the ascending phase of SC~25, $\langle Q_{\rm Fe}\rangle$ increased steadily from 10.25 (2020) through 11.15 (2023) and 11.37 (2024) to 11.65 (2025). The SC~25 values surpass the SC~24 maximum, consistent with SC~25 being more active. The coefficient $r$~=~0.86 is notably higher than reported for SC~23 ($r$~=~0.64; \citealt{Song+etal+2021}), suggesting that $\langle Q_{\rm Fe}\rangle$ tracks the solar cycle more consistently when both the ascending and descending phases are well sampled.

\textit{Elemental abundance ratio (Fe/O). ---} Panel~(d) of Figure~\ref{Fig2} displays the variation of Fe/O, -- a proxy for the FIP effect, which depends on both global magnetic activity and local processes such as wave–particle interactions and reconnection geometry. The coefficients ($r$ = 0.57, $\rho$ = 0.67, $\tau$ = 0.49) indicate a moderate, monotonic but nonlinear relationship with solar activity, with scatter suggesting additional modulation beyond SSN. Fe/O increased from 0.18 (2009) to 0.34 during SC~24 maximum (2014), decreased to 0.25 in 2018, and recovered during SC~25 to 0.33 (2024) and 0.35 (2025). The overall correlation with SSN is moderate ($r$~=~0.57), compared to $r$~=~0.83 for SC~23 \citep{Song+etal+2021}. A moderate correlation indicates that Fe/O responds not only to the global SSN trend but also to local conditions such as reconnection geometry and Alfv\'{e}n wave generation in CME source regions \citep{Laming+2015}.

\subsection{Correlations Between Charge States and Elemental Abundances}
\label{sect:results_corr}

\begin{figure}
\centering
\includegraphics[width=0.9\textwidth]{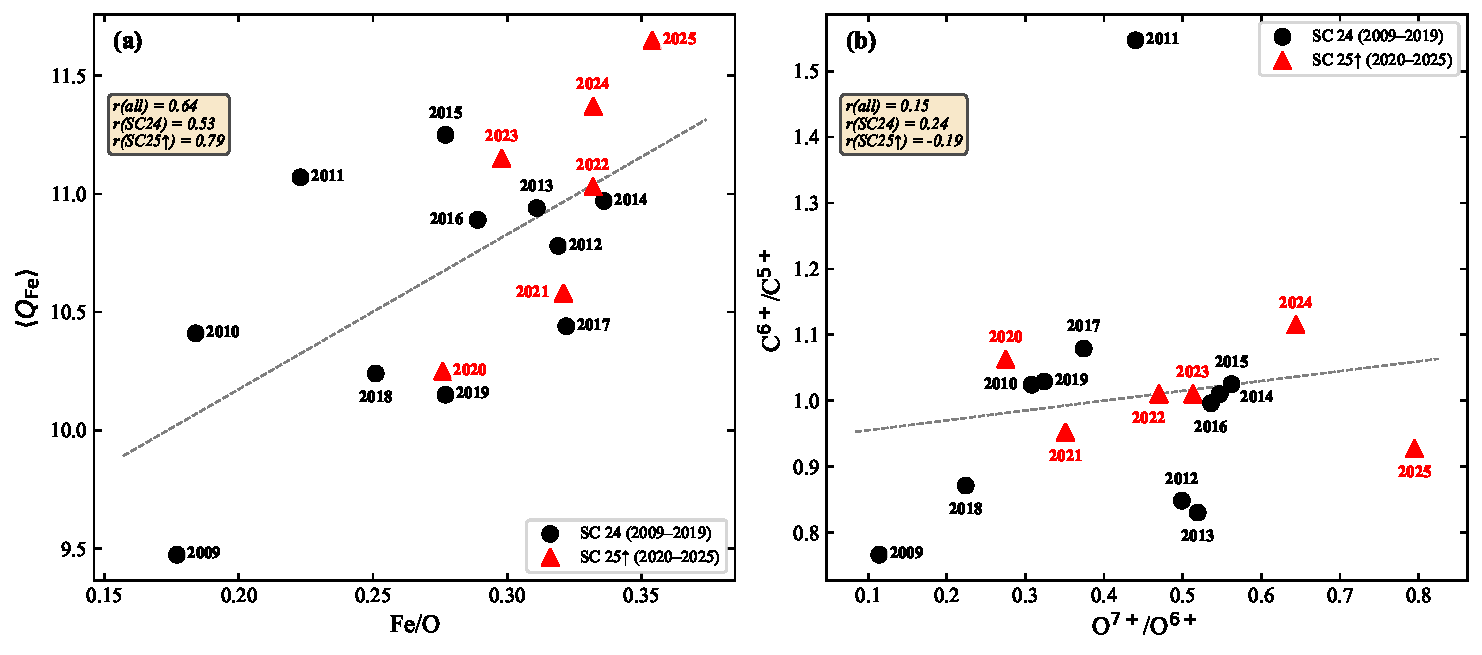}
\caption{Scatter plots of annual mean compositional parameters within ICMEs. (a)~$\langle Q_{\rm Fe}\rangle$ versus Fe/O; (b)~C$^{6+}$/C$^{5+}$ versus O$^{7+}$/O$^{6+}$. Black circles: SC~24 (2009--2019); red triangles: SC~25 (2020--2025). Year labels and correlation coefficients are shown.}
\label{Fig3}
\end{figure}

Figure~\ref{Fig3} presents scatter plots of the annual mean values. Panel~(a) of Figure~\ref{Fig3} shows the relationship between $\langle Q_{\rm Fe}\rangle$ and Fe/O, with black circles representing SC~24 (2009--2019) and red triangles representing SC~25 (2020--2025). The overall correlation is $r$~=~0.64 (strong), with SC~24 alone yielding $r$~=~0.53 (moderate) and the SC~25 data showing $r$~=~0.79 (strong). The stronger correlation for SC~25 suggests that during the ascending phase and maximum, coronal heating and elemental fractionation track each other more closely. For comparison, \citet{Song+etal+2021} reported $r$~=~0.76 between $\langle Q_{\rm Fe}\rangle$ and Fe/O for SC~23.

Panel~(b) of Figure~\ref{Fig3} displays the relationship between C$^{6+}$/C$^{5+}$ and O$^{7+}$/O$^{6+}$. The overall correlation is very weak ($r$~=~0.15), with SC~24 showing $r$~=~0.24 and SC~25 showing $r$~=~$-$0.19. The negative correlation in SC~25 arises because C$^{6+}$/C$^{5+}$ decreased from 1.12 (2024) to 0.93 (2025) while O$^{7+}$/O$^{6+}$ continued to rise to 0.80. This divergent behavior reflects the different freeze-in distances and temperature sensitivities of C and O charge states: C$^{6+}$/C$^{5+}$ is shaped predominantly by low-coronal conditions and is also subject to the SWICS~2.0 saturation truncation that suppresses its dynamic range, while O$^{7+}$/O$^{6+}$ responds to broader coronal temperature variations and is less affected by saturation.
 
The very weak overall correlation ($r$~=~0.15) between C$^{6+}$/C$^{5+}$ and O$^{7+}$/O$^{6+}$ is consistent with the SWICS~2.0 saturation effect on C$^{6+}$/C$^{5+}$: while O$^{7+}$/O$^{6+}$ retains its full dynamic range (0.11--0.80), the C$^{6+}$/C$^{5+}$ range is compressed to 0.83--1.12 in the post-anomaly period, eliminating the proportional relationship between the two ratios that would be expected from their common dependence on coronal electron temperature.

\subsection{Hot ICMEs}
\label{sect:results_hot}

\begin{figure}
\centering
\includegraphics[width=0.9\textwidth]{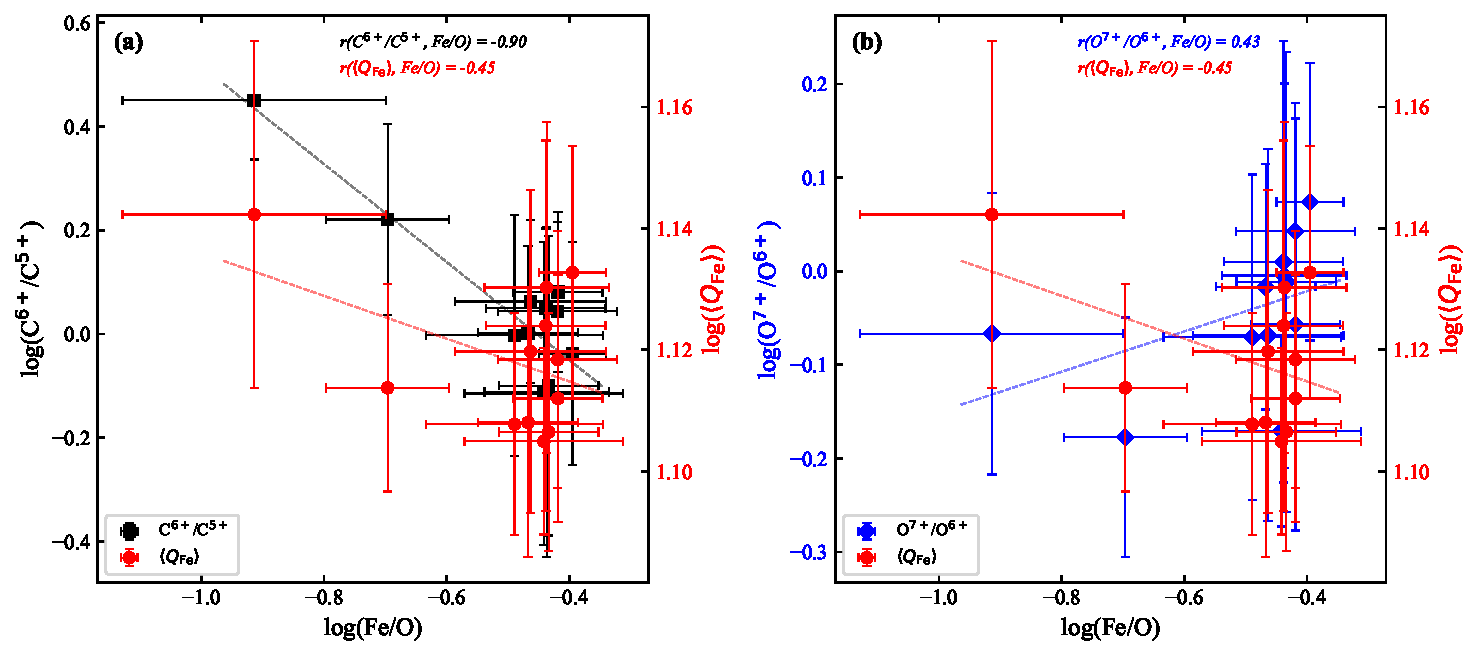}
\caption{Composition parameter relationships within hot ICMEs ($\langle Q_{\rm Fe}\rangle > 12$). (a)~Logarithmic C$^{6+}$/C$^{5+}$ (black squares, left axis) and $\langle Q_{\rm Fe}\rangle$ (red circles, right axis) versus logarithmic Fe/O. (b)~Logarithmic O$^{7+}$/O$^{6+}$ (blue diamonds, left axis) and $\langle Q_{\rm Fe}\rangle$ (red circles, right axis) versus logarithmic Fe/O. Pearson coefficients and best-fit lines are shown.}
\label{Fig4}
\end{figure}

Figure~\ref{Fig4} displays the relationships between composition parameters within hot ICMEs ($\langle Q_{\rm Fe}\rangle > 12$ for at least 6 hours; \citealt{Lepri+etal+2001}). Logarithmic axes are used because the hot ICME compositional parameters span a wide dynamic range, and logarithmic scaling better reveals power-law relationships between charge states and elemental abundances. This approach also reduces the visual dominance of extreme outliers and is consistent with previous studies of ICME composition correlations \citep{Zurbuchen+etal+2016}. Panel~(a) of Figure~\ref{Fig4} shows logarithmic C$^{6+}$/C$^{5+}$ (black squares) and $\langle Q_{\rm Fe}\rangle$ (red circles) versus logarithmic Fe/O. The C$^{6+}$/C$^{5+}$ ratio exhibits a strong anti-correlation with Fe/O ($r$~=~$-$0.90), while $\langle Q_{\rm Fe}\rangle$ shows a moderate negative correlation ($r$~=~$-$0.45). Panel~(b) of Figure~\ref{Fig4} shows O$^{7+}$/O$^{6+}$ (blue diamonds) and $\langle Q_{\rm Fe}\rangle$ versus Fe/O. O$^{7+}$/O$^{6+}$ displays a moderate positive correlation with Fe/O ($r$~=~0.43), while $\langle Q_{\rm Fe}\rangle$ maintains its negative correlation ($r$~=~$-$0.45).

The strong anti-correlation between C$^{6+}$/C$^{5+}$ and Fe/O in hot ICMEs ($r$~=~$-$0.90) implies that in the most Fe-enriched hot ICMEs, rapid ionization driven by flare reconnection suppresses lower carbon charge states. Conversely, the positive correlation between O$^{7+}$/O$^{6+}$ and Fe/O ($r$~=~0.43) supports the view that oxygen ions trace the high-temperature evolution of CME source regions, where both elevated temperatures and enhanced FIP fractionation occur during energetic flare-driven eruptions \citep{Henke+etal+2001, Lepri+etal+2013}.

\section{Discussion}
\label{sect:discussion}

\subsection{Comparison with Solar Cycle~23 Results}

Table~\ref{Tab2} summarizes the comparison of the Pearson coefficients between our study and \citet{Song+etal+2021}.

\begin{table}
\bc
\begin{minipage}[]{100mm}
\caption[]{Comparison of the Pearson Correlation Coefficients Between ICME Compositional Parameters and SSN for SC~23 and SC~24/25}\label{Tab2}
\end{minipage}
\setlength{\tabcolsep}{6pt}
\small
\begin{tabular}{lcc}
\toprule
Parameter & SC~23 (1998--2011) & SC~24/25 (2009--2025) \\
 & \citet{Song+etal+2021} & This study \\
\midrule
$\langle Q_{\rm Fe}\rangle$ & 0.64 & 0.86 \\
O$^{7+}$/O$^{6+}$ & 0.80 & 0.85 \\
C$^{6+}$/C$^{5+}$ & 0.63 & 0.17 \\
Fe/O & 0.83 & 0.57 \\
\bottomrule
\end{tabular}
\ec
\end{table}

The most notable differences are the substantially higher value of $r$ for $\langle Q_{\rm Fe}\rangle$ (0.86 versus 0.64) and the markedly lower $r$ for C$^{6+}$/C$^{5+}$ (0.17 versus 0.63). The enhanced correlation of $\langle Q_{\rm Fe}\rangle$ with SSN benefits from the extended coverage spanning both the complete SC~24 (ascending, maximum, and descending phases) and the rise of SC~25 through its maximum, providing better sampling of the full range of solar activity levels. The significantly reduced $r$ of C$^{6+}$/C$^{5+}$ (0.17 versus 0.63) is primarily an instrumental artifact caused by the SWICS~2.0 upper-end saturation truncation at C$^{6+}$/C$^{5+}$ $\geq$~1.486, which clips the high-end tail of the distribution and suppresses the solar maximum signal (see Section~\ref{sect:results_SC} for a detailed analysis). A secondary contribution comes from the fact that C$^{6+}$/C$^{5+}$ freezes-in at lower heliocentric distances ($<$1.5~$R_\odot$), making it more sensitive to local conditions in the low corona and to potential contamination from surrounding solar wind at imprecise ICME boundaries. The moderate $r$ for Fe/O (0.57 versus 0.83) indicates that elemental fractionation during the weaker SC~24 was influenced more strongly by local source region conditions relative to the global SSN trend.

\subsection{Physical Implications}

The solar cycle dependence of ICME composition is understood through coronal heating and CME eruption processes. During solar maximum, more CMEs originate from active regions with complex magnetic topologies, where magnetic reconnection is more energetic \citep{Song+etal+2016, Song+etal+2021, Li+etal+2023}. The higher electron temperatures lead to more elevated ionic charge states, explaining the strong correlations of $\langle Q_{\rm Fe}\rangle$ and O$^{7+}$/O$^{6+}$ with SSN. The Fe/O ratio reflects FIP-driven fractionation enhanced by Alfv\'{e}n waves generated through reconnection during solar maximum \citep{Laming+2015}. Its persistence as a positive (moderate) correlation across SC~24/25 confirms this mechanism operates across cycles of different strengths.

\subsection{Limitations}

Several limitations should be noted. The discontinuity between SWICS~1.1 and 2.0 data introduces some uncertainty. The number of ICMEs varies significantly between years (e.g., 2 events in 2025 with SWICS data versus 30 in 2015), affecting statistical robustness. Additionally, compositional data represent a single spacecraft trajectory through a three-dimensional structure with spatially varying properties.

\section{Conclusions}
\label{sect:conclusion}

This study statistically examined the relationship between ICME compositional signatures and solar activity during 2009--2025, covering SC~24 and SC~25 through its maximum. The agreement between Pearson and Spearman coefficients suggests that the relationship is both linear and monotonic, while the Kendall coefficient confirms its robustness against outliers and small sample size. Using SWICS data from the ACE spacecraft, we analyzed $\langle Q_{\rm Fe}\rangle$, C$^{6+}$/C$^{5+}$, O$^{7+}$/O$^{6+}$, and Fe/O for 307 ICMEs. Our main findings are:

\begin{enumerate}
\item All composition parameters show solar cycle dependence, increasing with SSN during the ascending phases and decreasing during descending phases.

\item $\langle Q_{\rm Fe}\rangle$ and O$^{7+}$/O$^{6+}$ show strong monotonic correlations with SSN, confirming enhanced coronal heating at solar maximum. In contrast, C$^{6+}$/C$^{5+}$ shows no significant relationship ($r$~=~0.17, $\rho$~=~0.01), which is primarily caused by the SWICS~2.0 upper-end saturation truncation (C$^{6+}$/C$^{5+}$ $\geq$~1.486 removed) that clips the solar maximum signal, rather than indicating a physical absence of solar cycle dependence. Fe/O exhibits a weaker, nonlinear dependence ($r$~=~0.57), reflecting additional local influences from local plasma conditions.

\item Compared with SC~23, the Pearson coefficient ($r$) increased from 0.64 to 0.86 for $\langle Q_{\rm Fe}\rangle$ and from 0.80 to 0.85 for O$^{7+}$/O$^{6+}$, while it decreased from 0.63 to 0.17 for C$^{6+}$/C$^{5+}$ and from 0.83 to 0.57 for Fe/O.

\item In hot ICMEs ($\langle Q_{\rm Fe}\rangle > 12$), Fe/O is strongly anti-correlated with C$^{6+}$/C$^{5+}$ ($r$~=~$-$0.90) and moderately anti-correlated with $\langle Q_{\rm Fe}\rangle$ ($r$~=~$-$0.45), but positively correlated with O$^{7+}$/O$^{6+}$ ($r$~=~0.43).

\item SC~25 exhibits higher $\langle Q_{\rm Fe}\rangle$ (up to 11.65) and O$^{7+}$/O$^{6+}$ (up to 0.80) than SC~24 maximum values, consistent with SC~25 being a more active cycle.
\end{enumerate}

Overall, the correlations reveal a hierarchy in ICME compositional responses to solar activity: $\langle Q_{\rm Fe}\rangle$ and O$^{7+}$/O$^{6+}$ show strong monotonic trends, Fe/O a weaker nonlinear dependence, and C$^{6+}$/C$^{5+}$ no monotonic relationship, reflecting sensitivity to local rather than global conditions. These findings reinforce that ICME composition is a reliable proxy for tracing coronal temperature and energetic eruptions across solar cycles. Future research should integrate multi-spacecraft observations to further constrain coronal source properties and refine models of CME-to-ICME evolution.

\begin{acknowledgements}
The authors thank Dr. H. Song for the motivation to write this article. We thank Prof. Nandita Srivastava for fruitful discussion. We acknowledge the use of data from CDAWeb and WDC-SILSO Royal Observatory of Belgium. We also thank Drs. M. Raines, L. Zhao, S. Lepri, and the ACE team. This work is supported by India-Uzbekistan joint project INT-UZBEK/P-15 and UZB-IND-2021-95. We are grateful to an anonymous referee for valuable suggestions that led to significant improvements and additions to the manuscript.
    \end{acknowledgements}

\bibliographystyle{raa}
\bibliography{bibtex}

\end{document}